\documentclass[%
 reprint,
superscriptaddress,
groupedaddress,
 amsmath,amssymb,
 aps,
pra,
]{revtex4-2}
\usepackage{physics}
\usepackage{hyperref}
\usepackage{graphicx}
\usepackage{dcolumn}
\usepackage{bm}
\usepackage{capt-of}
\newcommand{\W}{\mathrm{W}^\prime}

\begin{document}

\preprint{APS/123-QED}

\title{Probing a Quarkophobic ${\mathbf{W}}^\prime$ at the High-Luminosity LHC via Vector Boson Fusion and Lorentz-Equivariant Point Cloud Learning}

\author{U. S. Qureshi}%
 \email{uqureshi@cern.ch}
\author{A. Gurrola}
 \email{alfredo.gurrola@vanderbilt.edu}
 
\affiliation{Department of Physics and Astronomy,
Vanderbilt University, Nashville, TN, U.S.A.}%

\author{J. D. Ruiz-\'{A}lvarez}
  \email{josed.ruiz@udea.edu.co}
\affiliation{
Instituto de F\'isica, Universidad de Antioquia, A.A. 1226 Medell\'in, Colombia.
}%


\date{\today}

\begin{abstract}
The addition of a heavy charged vector gauge boson $\W$ to the Standard Model (SM) with negligible quark couplings (``quarkophobic'') and triple gauge couplings can address issues with the SM, such as the B-meson anomalies and recent discrepancies in the W boson mass measurements. We present a phenomenology study probing $\W$ production through weak boson fusion in proton-proton collisions at the Large Hadron Collider. We operate under a simplified model with a large $\W$ decay width and consider final states with two jets, large missing transverse momentum, and one light lepton. Notably, we use point cloud learning for the first time in a BSM search---specifically, a novel Lorentz-Equivariant Geometric Algebra Transformer---providing significant improvement in signal sensitivity compared to traditional methods.
\end{abstract}

\maketitle


The Standard Model (SM) of particle physics has been extensively tested at CERN's Large Hadron Collider (LHC) as a theory describing elementary particles and their interactions below the \textrm{TeV}-scale. However, as experiments probe higher energies, observations have emerged that contradict the SM. These include the origins of dark matter \cite{WMAP:2012nax, Planck:2018vyg, Bertone:2004pz}, electroweak symmetry breaking scales \cite{Branco:2011iw, Gori:2016zto, BhupalDev:2014bir, Liu:2023jbq}, baryon asymmetry \cite{Sakharov:1967dj, Dine:2003ax}, neutrino masses \cite{Kajita:2016cak}, the anomalous muon magnetic moment \cite{Muong-2:2023cdq, Muong-2:2024hpx}, discrepancies in the $R_{(D)}$ and $R_{(D^{*})}$ ratios from B-meson decays \cite{LHCb:2023rd, BaBar:2013mob, Belle:2015qfa, Belle:2016ure, Belle:2016dyj}, and potential inconsistencies in the W boson mass measurements \cite{ParticleDataGroup:2022pth, Erler_2019, Awramik_2004, CDF:2022hxs, cmscollaboration2024highprecisionmeasurementwboson}. As a result, the SM is widely seen as a low-energy effective theory, increasing the need for beyond the SM (BSM) physics.

\begin{figure}
    \centering
    \begin{minipage}{.49\linewidth}
  \centering
  \includegraphics[width=\linewidth]{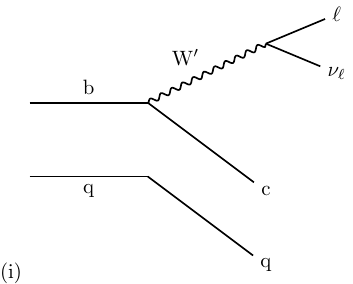}
\end{minipage}%
\begin{minipage}{.49\linewidth}
  \centering
  \includegraphics[width=\linewidth]{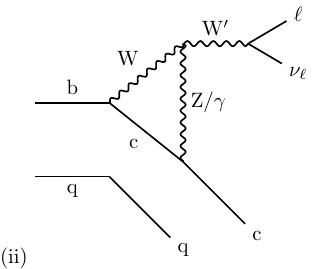}
\end{minipage}%

    \begin{minipage}{.49\linewidth}
  \centering
  \includegraphics[width=\linewidth]{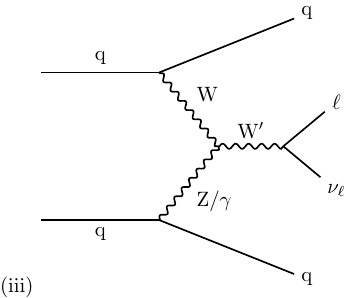}
\end{minipage}%
\begin{minipage}{.49\linewidth}
  \centering
  \includegraphics[width=\linewidth]{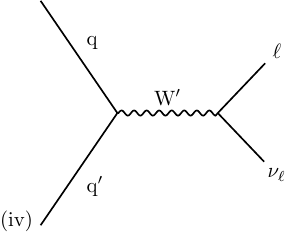}
\end{minipage}%
\caption{Representative diagrams for: (i) a traditional $\W$ mediating a process contributing to the b-anomalies; (ii) a quarkophobic $\W$ with TGC mediating a process involved in the b-anomalies; (iii) VBF $\W$ production and subsequent decay to a lepton and a neutrino; (iv) Drell-Yan $\W$ production and subsequent decay to a lepton and a neutrino.}
\label{fig:feynmanns}
\end{figure}

The production of a heavy, charged resonance $\W$, which could mediate processes that contribute to B-meson anomalies, has been a subject of interest for both theory and experiment \cite{PhysRevD.11.566, PhysRevD.12.1502, Appelquist_2001, Appelquist:2000nn, Agashe:2008jb}. In addition, if one believes the W boson mass is inconsistent with the SM predictions, the presence of a heavy $\W$ could account for the observed mass discrepancy \cite{CDF:2022hxs}. Even if the W boson mass aligns with SM expectations, investigating a $\W$ boson is crucial, especially in previously uncharted territory, to uncover signs of new physics. While the nature, mass, couplings, and quantum numbers of this hypothetical particle remain undetermined, current experimental results from the LHC \cite{ATLAS:2014txl, ATLAS:2018uca, ATLAS:2023ibb, CMS:2023gte, CMS:2022tdo, CMS:2012nze} have imposed constraints assuming the simplest BSM scenarios. Traditional searches consider a $\W$ mediator that couples to both quarks and leptons of the SM. Figure \ref{fig:feynmanns} (i) shows a representative Feynman diagram for this scenario that can address the B-meson anomalies, while Figure \ref{fig:feynmanns} (iv) shows a typical production mode to probe this at the LHC, which assumes a narrow $\W$ width $\Gamma / m(\W) \sim \mathcal{O}(<10\%)$.

In this Letter, we present a new study on the production of a $\W$ boson at the LHC with negligible couplings to the SM quarks (quarkophobic). This $\W$ couples to the SM weak bosons through triple gauge couplings (TGC), which could also explain the B-meson anomalies. The diagram for such a model is depicted in Fig.~\ref{fig:feynmanns} (ii). We note the CMS search in Ref.~\cite{CMS:VBFWprimeHVT} examines $\W$ decays to SM weak bosons within Heavy Vector Triplet (HVT) models but lacks sensitivity when fermion couplings are suppressed. These searches also assume a narrow decay width, as their strategies rely on this due to experimental resolution limitations. In contrast, this study presents a novel $\W$ model with a significantly larger decay width, which has not been explored before.

Such a scenario naturally motivates a probe of the $\W$ via vector boson fusion (VBF) production from high-energy proton-proton collisions at the LHC. This paves the way for a search with a single high-energy lepton from the $\W$ decay, missing momentum ($p^{\mu}_{\textrm{miss}}$), and two VBF-tagged jets as shown in Fig.~\ref{fig:feynmanns} (iii). The VBF topology has proved to be a powerful experimental tool for searches at the LHC due to its remarkable control over SM backgrounds while also creating a kinematically boosted topology \cite{CMS:2016ucr, Florez:2016uob, CMS:2015jsu, VBFGraviton, VBFADM, VBFHiggsino, VBFHN, PhysRevD.92.095009, Dutta_2013, Delannoy:2013ata, Dutta:2014jda, Dutta:2013gga, Florez:2021zoo, Avila:2018sja, Arnowitt:2008bz, VBFHiggsino, Qureshi:2024cmg}. 

We develop a novel search strategy using the cutting-edge Lorentz-Equivariant Geometric Algebra Transformer (L-GATr) architecture---designed specifically for high-energy physics. L-GATr operates on particle four-momentum point clouds, incorporating Lorentz equivariance through a geometric algebra framework, allowing it to leverage physical symmetries to learn accurate representations. The transformer-based architecture also enables scalability and generalization to large datasets. This approach significantly outperforms traditional methods like boosted decision trees (BDTs) and multi-layer perceptrons (MLPs), which lack these physical inductive biases. The L-GATr classifier's outputs are used in a profile-binned likelihood fit to assess signal significance. This is the first use of L-GATr (and point clouds) in an LHC phenomenological analysis, demonstrating its potential to redefine how machine learning can be utilized for new physics searches.

The rest of this Letter is organized as follows: we begin by summarizing the minimal $\W$ model, followed by a description of the Monte Carlo simulation samples. Next, we outline our machine learning workflow and present the main results, concluding with a brief discussion.

The minimal $\W$ model extends the SM by introducing a new $\W$ particle with minimal couplings to SM particles, as defined in Eqs.~\ref{eq:weakbosons1}, \ref{eq:weakbosons2}, and \ref{eq:lep1}. The interactions with SM electroweak bosons are given by:
\begin{equation}
  \label{eq:weakbosons1}
  \begin{aligned}
      \mathcal{L}^{1}_{VWW^\prime} = ~&g_1^V V^\mu \big(W_\mu^{-} W_{\nu}^{'+} - W_\mu^{+} W_{\nu}^{'-} \\ &+ W_\mu^{'-} W_{\nu}^{+} - W_\mu^{'+} W_{\nu}^{-} \big).
  \end{aligned}
\end{equation}
These are complemented by higher-order terms, given by:
\begin{equation}
  \label{eq:weakbosons2}
  \mathcal{L}^2_{VWW'} = g_2^V \big(W_\mu^{+} W_\nu^{'-} V^{\mu \nu} + W_\mu^{'+} W_\nu^{-} V^{\mu \nu} \big).
\end{equation}
The VBF $\W$ cross section is governed by the coupling $g_{1,2}^V$. 
Interactions with the SM leptons are described by
\begin{equation}
     \mathcal{L}_\ell = \sum_\ell \overline{\nu}_\ell \gamma_\mu \big(g_\ell^R (1 + \gamma^5) + g_\ell^L (1 - \gamma^5)\big) W^{'\mu} \ell,
        \label{eq:lep1}
\end{equation}
where $V_{\mu\nu} = \partial_\mu V_\nu - \partial_\nu V_\mu$, $\ell = \{e,\mu, \tau\}$, $\nu_\ell = \{\nu_e, \nu_\mu, \nu_\tau \}$, $V=\{\mathrm{Z}, \gamma\}$, and $g_{\ell}^{L}$ ($g_{\ell}^{R}$) represent the left-handed (right-handed) $\W$ couplings to leptons. In our studies we set $g_{\ell}^{L}=g_{\ell}^{R}$, and henceforth refer to these coupligs as $g_{\ell}^{L, R}$. The $\W \to \nu \ell$ decay vertex and branching fraction are governed by $g_{\ell}^{L, R}$ and $g_{1,2}^V$. For $V=\mathrm{Z}$ in Eqs.~\ref{eq:weakbosons1}--\ref{eq:lep1}, the leading order contribution to the $\W$ decay width is $\frac{(g_{1}^V)^{2}m(W')^{5}}{192\pi m_{W}^{2}m_{Z}^{2}}$, where $m_{W}$ and $m_{Z}$ are the masses of the SM W and Z bosons. The decay width increases rapidly with $m(\W)$, surpassing $m(\W)$ for masses above about 500 GeV when $g_{1}^V = 1$. For $V=\gamma$, the leading order width scales as $\frac{(g_{1}^V)^{2}m(W')^{3}}{m_{W}^{2}}$, increasing less rapidly with $m(\W)$. In both cases $V=\{\mathrm{Z}, \gamma\}$, the maximal coupling $g_{\textrm{max}}^{V}$ is defined as the coupling value $g_{1}^V = g_{2}^V$, for a given $m(\W)$, where the width equals the mass. The model is valid when the coupling is less than or equal to this value. 

For this study, we set the couplings $g_{i}^{V}$ equal to $g_{\textrm{max}}^{V}$, and adopt an effective framework similar to the simplified models often used in BSM searches at the LHC: an implementation that simulates relevant interactions, specifically the production of a $\W$ boson via the VBF process at the LHC, followed by its decay into SM leptons. 


The minimal $\W$ model is implemented with \textsc{FeynRules} \cite{Alloul:2013bka}, and the resulting UFO \cite{Degrande:2011ua} is used to generate Monte Carlo samples. Parton-level events are produced with \textsc{MadGraph5\_aMC@NLO 2.9.22} \cite{Alwall:2014hca, Frederix:2018nkq} at next-to-leading order in QCD for pp collisions at $\sqrt{s} = 13.6$ TeV, using the \textsc{NNPDF3.0\_NLO} \cite{NNPDF:2014otw} parton distribution functions. These events are interfaced with \textsc{Pythia 8.2.30} \cite{Sjostrand:2014zea, Sjostrand:2007gs} for parton showering and hadronization. \textsc{Delphes 3.4.1} \cite{deFavereau:2013fsa} simulates detector effects using CMS parameters with 140 pileup interactions. Jets are clustered using the anti-$k_t$ algorithm~\cite{Cacciari_2008} with a distance parameter $R = 0.4$ using \textsc{FastJet} 3.4.2~\cite{Cacciari_2012}. The MLM algorithm is used for jet matching and merging, with \texttt{xqcut} and \texttt{qcut} set to 30 and 45, respectively, ensuring smooth differential jet rates.
\begin{figure}
    \centering
    \includegraphics[width=\linewidth]{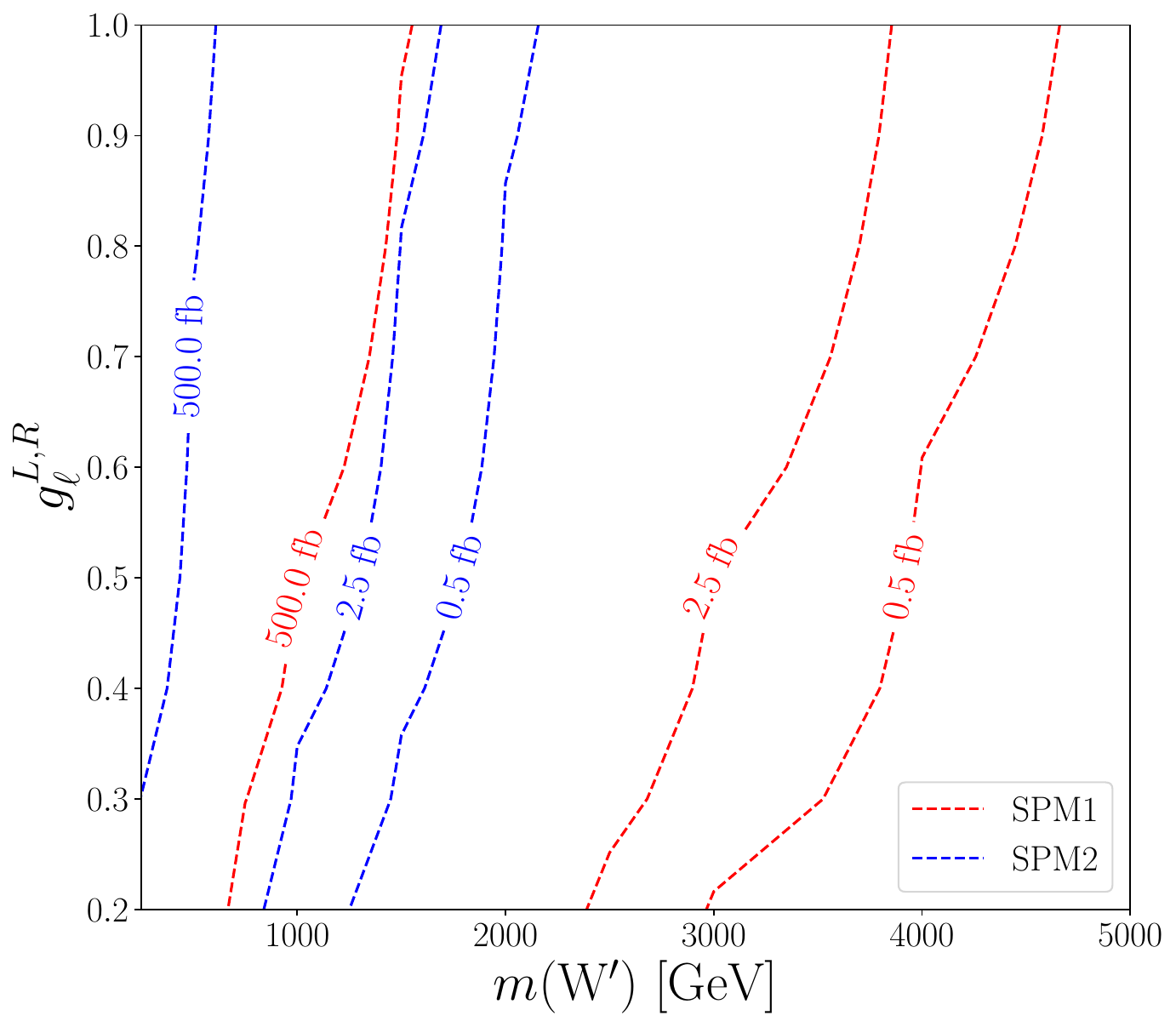}
    \caption{The $\textrm{pp} \to \W j j, (\W \to \nu \ell)$ production cross-section contours in femtobarns, as a function of the $\W$ mass and the lepton coupling $g_\ell^{L,R}$.}
    \label{fig:xsection}
\end{figure}

\begin{figure}
    \centering
    \includegraphics[width=\linewidth]{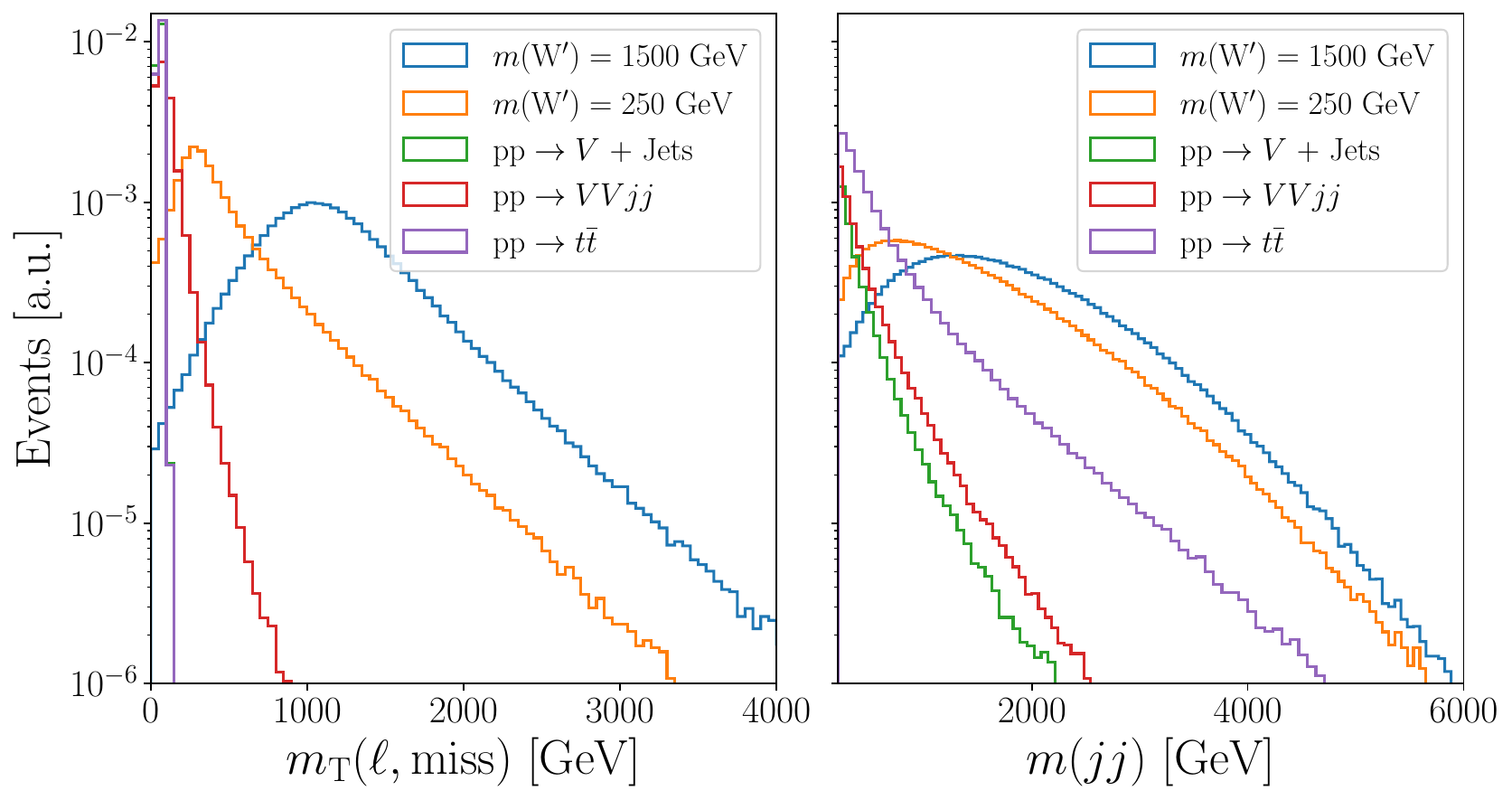}
    \caption{Reconstructed transverse mass (left) and dijet invariant mass (right) distributions for benchmark SPM1 signal scenarios and dominant SM backgrounds.}
    \label{fig:kinematics}
\end{figure}

Signal samples are produced considering pure electroweak single $\W$ production with two associated jets ($\textrm{pp} \to \W j j$) followed by $\W \to \nu \ell$ decays where $\ell =\{e,\mu\}$. We consider two sets of signal samples: (i) $V=\gamma$, with $g_{1,2}^{\gamma}=g_{\textrm{max}}^{\gamma}$ and $g_{1,2}^{\mathrm{Z}}=0$; (ii) $V=\mathrm{Z}$, with $g_{1,2}^{\mathrm{Z}}=g_{\textrm{max}}^{\mathrm{Z}}$ and $g_{1,2}^{\gamma}=0$. We refer to the former as simplified phenomenological model 1 (SPM1), and the latter SPM2. 
Fig.~\ref{fig:xsection} shows contours of constant value for the product of the cross section and branching fraction, as a function of $m(\W)$ and $g_{\ell}^{L,R}$ for both scenarios. All signal cross sections are obtained with the following parton-level cuts: $|\eta(\ell)| < 2.7$, $p_{\mathrm{T}}(j) > 10$ GeV, $\abs{\Delta \eta(jj)} > 2.5$, and $m(jj)>200$ GeV. The $\abs{\Delta \eta(jj)}$ and $m(jj)$ cuts suppress non-VBF contributions. For SPM1 (SPM2) with $g_\ell^{L,R} = 1$, the cross section is 115 (5.81) pb at $m(\W) = 250$ GeV and decreases to 0.0083 (0.0024) pb at $m(\W) = 3$ (1.5) TeV.  

Since we focus on final states with two VBF-tagged jets, missing momentum, and one light lepton (electron or muon), the main SM background sources are: (i) top quark pair production (${t}\overline{{t}}$); (ii) single Higgs production (gluon-gluon fusion, VBF, and associated production); (iii) $V=\{\mathrm{Z},\mathrm{W}\}$ boson with jets; (iv) pure QCD multijet production; (v) diboson $VV=\{\mathrm{WW}, \mathrm{ZZ}, \mathrm{WZ}\}$; and (vi) triboson events ($\{VVV\}$). The dominant backgrounds are ${t}\overline{{t}}$, $V$ + jets, and $VV$ with jets, while the other processes contribute less than 1\% of the total background

Identifying leptons, light-quark or gluon jets, and bottom quarks is crucial for both signal identification and SM background reduction, enhancing discovery potential at the High-Luminosity LHC (HL-LHC). However, this may be challenging due to significant pileup. The impact of pileup on VBF processes and the need for mitigation at CMS and ATLAS is discussed in Ref.~\cite{CMS-PAS-FTR-13-014}. While the performance of upgraded detectors at the HL-LHC is beyond the scope of this work, we conservatively assume some degradation in identification efficiencies, using Ref.~\cite{CMS-PAS-FTR-13-014} as a benchmark with 140 average pileup interactions. 

The tracking efficiency for charged hadrons, affecting jet clustering and missing transverse momentum, is 97\% for $1.5 < |\eta| < 2.5$, dropping to 85\% at $|\eta| = 2.5$. For light leptons with $p_{\mathrm{T}} > 5$ GeV and $|\eta|< 1.5$, the identification efficiency is 95\%, with a misidentification rate of 0.3\% \cite{CMS-PAS-FTR-13-014,CMS_MUON_17001}. This efficiency decreases linearly with $\eta$ for $1.5 < |\eta| < 2.5$, reaching 65\% and 0.5\% misidentification at $|\eta| = 2.5$. These inefficiencies affect lepton kinematics. Following Ref.~\cite{CMSbtag}, we use the ``Loose'' working point of the DeepCSV algorithm \cite{Bols_2020}, which provides 70-80\% b-tagging efficiency and 10\% light quark misidentification.

The analysis of signal and background events is done utilizing machine learning event classifiers. Unlike conventional methods, machine learning models simultaneously consider all kinematic variables, allowing them to efficiently navigate the complex high-dimensional space of event kinematics. Consequently, machine learning models can enact sophisticated selection criteria that take into account the entirety of this space, making them ideal for high-energy physics applications \cite{hepmllivingreview}. 

Recently, significant effort has been dedicated to developing point-cloud-based machine learning models to process unordered sets of points with varying cardinalities and incorporate geometric invariances \cite{8099499, qi2017pointnetdeephierarchicalfeature, wang2019dynamicgraphcnnlearning, Guo_2021, zhao2021pointtransformer, zaheer2018deepsets, lee2019settransformerframeworkattentionbased, kothapalli2024equivariantvsinvariantlayers}. These models have found widespread applications in high energy physics, especially in jet classification \cite{Onyisi_2023, shen2023hierarchicalhighpointenergyflow, Athanasakos_2024, gambhir2024momentsclaritystreamlininglatent, Odagiu_2024, hammad2024multiscalecrossattentiontransformerencoder, qu2024particletransformerjettagging, Mikuni_2021, Komiske_2019, Qu_2020, Mikuni_2020, Mikuni_2021}. Notably, however, these models have not yet been utilized to probe BSM physics in the context of an LHC search.

For this study, we use the state-of-the-art L-GATr
architecture \cite{Brehmer:2024yqw, Spinner:2024hjm, brehmer2023geometric}. L-GATr is a deep learning model designed for high-energy data, which combines the strengths of a Transformer with a geometric algebra framework. Unlike traditional models such as BDTs and MLPs, L-GATr is designed with strong inductive biases tailored for high-energy physics data. Its learned representations are generated by acting on particle four-momenta and extended to higher orders within a geometric algebra framework. The model's layers are equivariant under Lorentz symmetry by construction. L-GATr adopts the Transformer architecture, enabling it to efficiently scale to large capacities and handle extensive input tokens.

The effectiveness of machine learning models such as BDTs, MLPs, and Transformers has been validated in numerous studies \cite{Ai:2022qvs, ATLAS:2017fak, Biswas:2018snp,  Chung:2020ysf, ttZprime, Chigusa:2022svv,  Florez2023, Arganda2024, flórez2024probinglightscalarsvectorlike,U1T3R_GammaGamma, Qureshi:2024ceh}. In Ref.~\cite{Onyisi_2023}, the authors demonstrate that point cloud strategies based on deep sets and edge convolutions outperform traditional methods like BDTs and MLPs for collider event classification. Building on this, Ref.~\cite{Brehmer:2024yqw} highlights that L-GATr outperforms the aforementioned baselines in a range of tasks across the LHC discovery stack, including regression, classification, and generative modeling, thus motivating its use in our study. We found that this architecture provides a substantial improvement in sensitivity, extending 95\% exclusion bounds on $m(\W)$ by 1.7 times compared to BDTs.
\begin{figure}
    \centering
    \includegraphics[width=\linewidth]{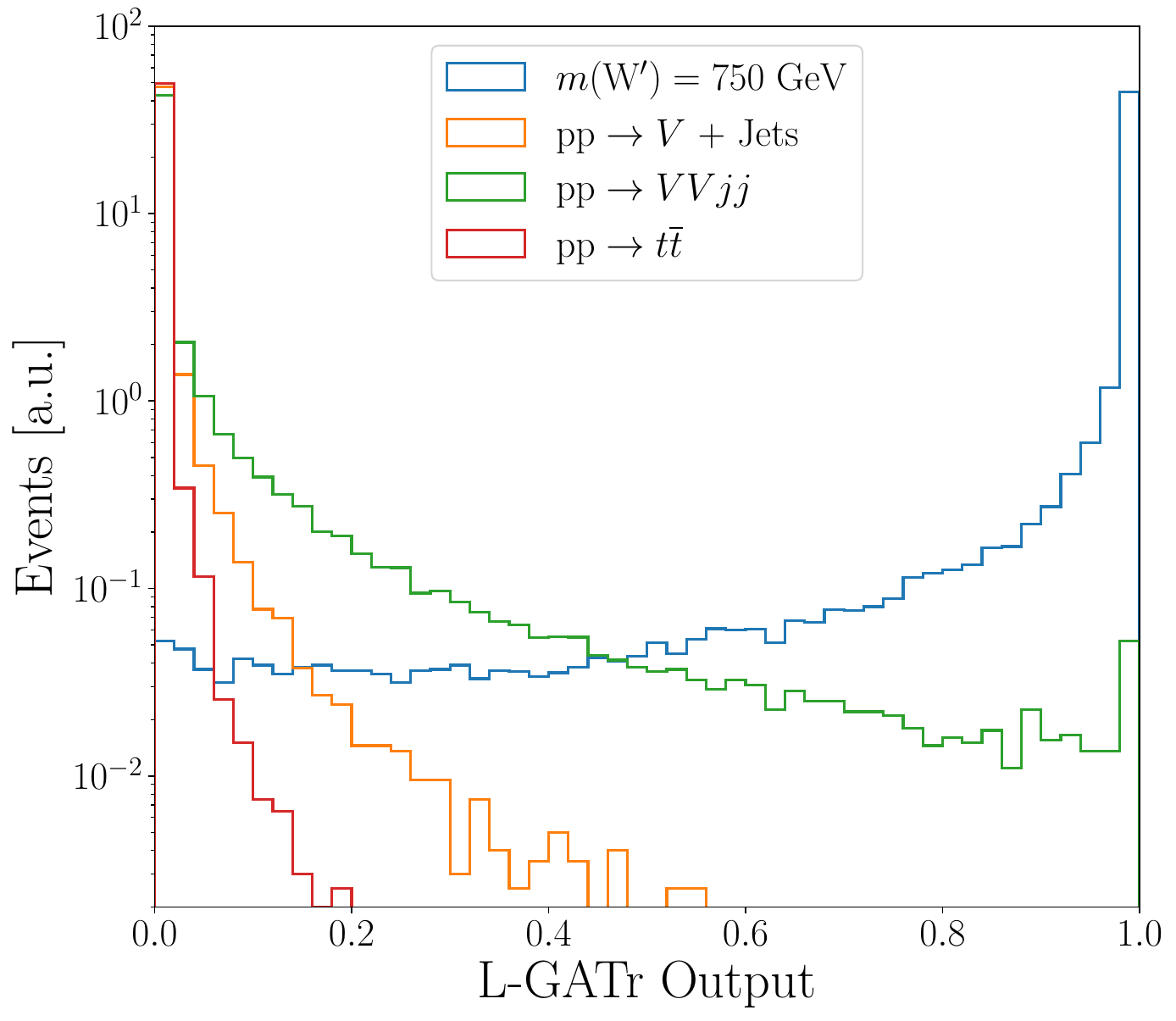}
    \caption{L-GATr output distributions for a $m(\W)=750$ GeV SPM2 signal and dominant backgrounds.}
    \label{fig:LGATROut}
\end{figure}

\begin{figure*}
    \centering
    \begin{minipage}{.49\textwidth}
  \centering

  \includegraphics[width=\linewidth]{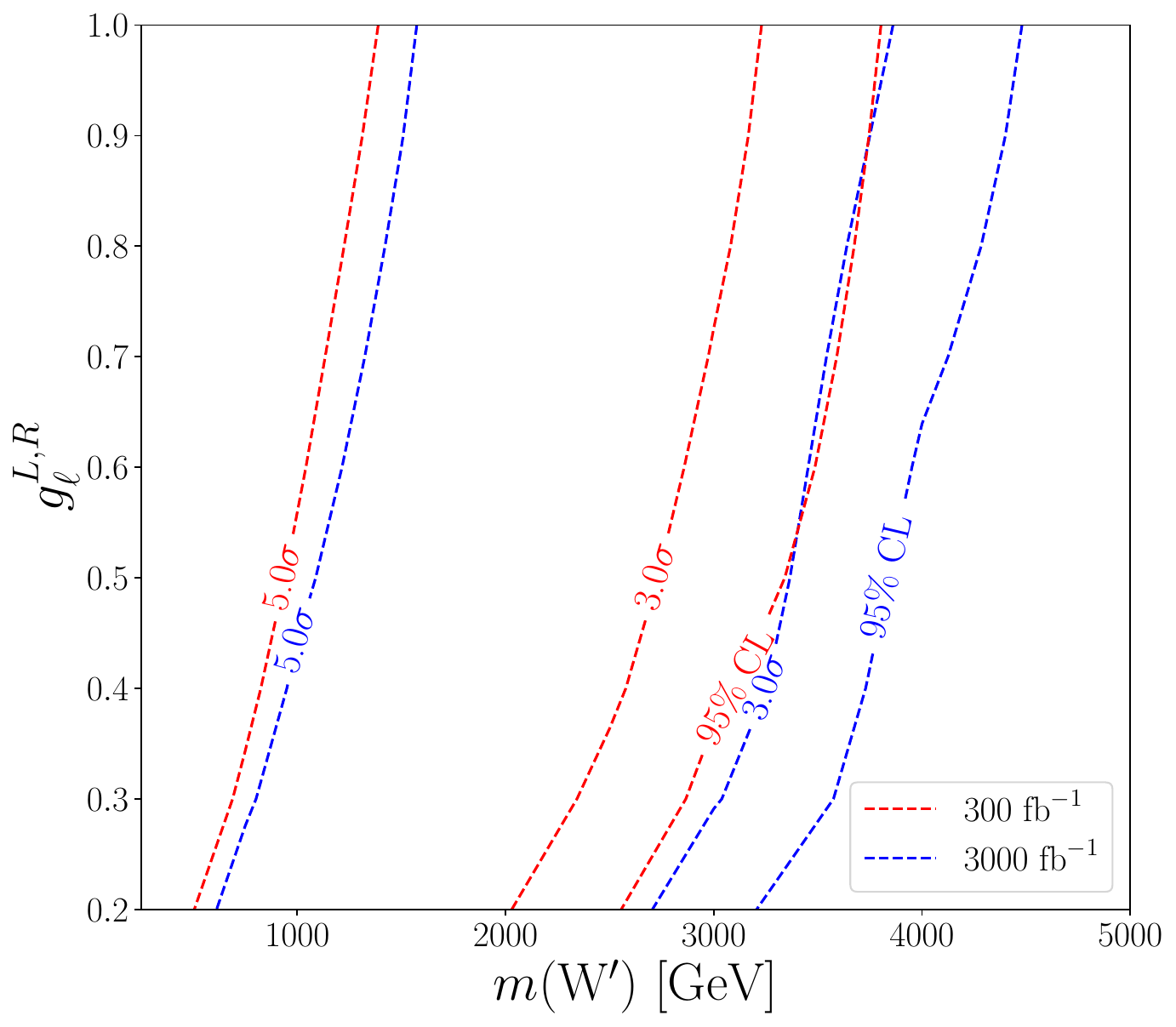}
\end{minipage}%
\begin{minipage}{.49\textwidth}
  \centering
  \includegraphics[width=\linewidth]{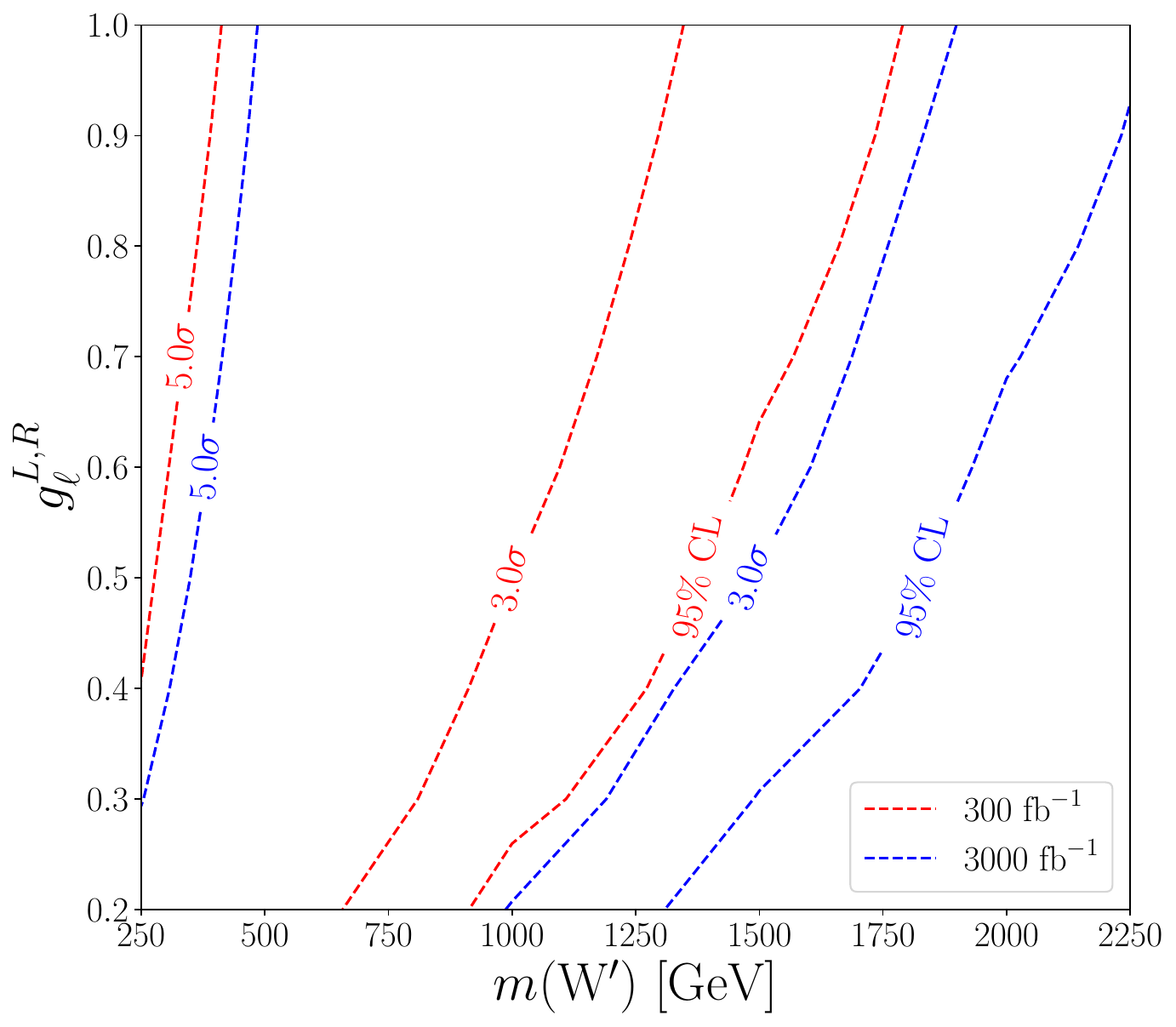}
\end{minipage}%
\caption{Projected signal significance contours as a function of the $\W$ mass and the lepton coupling $g_\ell^{L,R}$ for SMP1 (left) and SPM2 (right) signal scenarios for integrated luminosities of 300 $\mathrm{fb}^{-1}$ corresponding to the current end-of-run estimates (red) and 3000 $\mathrm{fb}^{-1}$ corresponding to the HL-LHC (blue). The regions inside the contours lie within discovery/exclusion potential.}
\label{fig:SS}
\end{figure*}

Our workflow uses a specialized \textsc{MadAnalysis5} script \cite{CONTE2013222} to extract particle kinematic data into a structured CSV format for machine learning. Cross section weighting is applied to balance the significance of signal and background events. The weighted datasets are filtered and then written to the CSV file for use by the machine learning algorithm. Signal and background events are modeled as point clouds, with each event represented by the four momenta of analysis level objects: $\{p^\mu(j_1), p^\mu(j_2), p^\mu(\ell), p^{\mu}_{\textrm{miss}}\}$. Although the four-momenta of the analysis-level objects serve as the inputs to the machine learning algorithm, Fig.~\ref{fig:kinematics} displays representative kinematic distributions derived from these 4-momentum vectors. These distributions are presented to emphasize the discriminating power of the VBF $\W$ production mechanism and to illustrate the information encoded within the 4-momentum vectors.

Training and evaluation of our ML models were performed on an Nvidia A100 GPU using the official \textsc{PyTorch} \cite{paszke2019pytorchimperativestylehighperformance} implementation of the \textsc{TaggingGATrWrapper}. The Lion optimizer \cite{chen2023symbolicdiscoveryoptimizationalgorithms} is combined with binary cross-entropy loss and a cosine annealing scheduler \cite{loshchilov2017sgdrstochasticgradientdescent}, with an initial learning rate of $0.0005$ and a weight decay of 0.01. The model has 32 hidden scalar channels, 16 multivector channels, 8 attention heads, and 6 blocks. A 90-10 train-test split and a batch size of $256$ are used, with the output on the test set for signal significance calculation.

For the filtering criteria, we require one identified lepton with $p_{\mathrm{T}} > 25$ GeV and $|\eta| < 2.4$. A veto is applied for events containing a well-identified $\tau_{\textrm{h}}$ (b-jet) with $p_{\mathrm{T}} > 20$ ($30$) GeV and $|\eta| < 2.5$. The VBF signal topology requires two forward jets in opposite hemispheres with large dijet invariant mass, while jets in background events are mostly central with small dijet masses. VBF criteria are applied by requiring at least two jets with $p_{\mathrm{T}} > 25$ GeV and $|\eta| < 5$. Jets with $|\Delta\eta(jj)| > 3.0$ and $\eta(j_{1}) \cdot \eta(j_{2}) < 0$ are combined to form VBF dijet candidates, which must satisfy $m(jj) > 250$ GeV.

Events passing the pre-selections are used as input for the machine learning algorithm, classifying them as signal or background. The L-GATr output ranges from 0 to 1, indicating the likelihood of an event being signal-like (near 1) or background-like (near 0). The full spectrum of the output as shown in Fig.~\ref{fig:LGATROut} is used in a profile-binned likelihood fit to estimate signal significance, considering two luminosity scenarios, $\mathcal{L}_{\mathrm{int}} = 300$ and 3000 $\mathrm{fb^{-1}}$, for the current and HL-LHC. For significance calculations, L-GATr output histograms in Fig.~\ref{fig:LGATROut} are normalized to $N = \mathcal{L}_\text{int}\cdot \sigma \cdot \epsilon$ where $\epsilon$ accounts for selection and reconstruction efficiencies. Significance is derived from the bin-by-bin yields in a profile likelihood fit, with $Z_\text{sig}$ calculated using the local $p$-value, similar to Refs.~\cite{ttZprime, Florez2023, flórez2024probinglightscalarsvectorlike,U1T3R_GammaGamma, Qureshi:2024ceh}.

Systematic uncertainties from experimental and theoretical considerations are incorporated as nuisance parameters. A 1-5\% uncertainty is applied for the choice of Parton Distribution Function (PDF) set, following PDF4LHC recommendations~\cite{Butterworth:2015oua}. This uncertainty slightly affects event yields but does not significantly impact the L-GATr output distribution's shape. PDF uncertainties are uncorrelated across signal and background processes, but correlated within bins for each process. Theoretical uncertainties due to missing higher-order contributions are estimated by varying the renormalization and factorization scales by a factor of 2, yielding uncertainties of 2-5\%, depending on the specific L-GATr bins. Experimental uncertainties include a 3\% luminosity uncertainty, treated as fully correlated across processes and bins, based on CMS measurements~\cite{lumiRef}. For light leptons, a 2\% uncertainty for reconstruction and identification and a 3\% uncertainty for momentum and energy estimation are included~\cite{CMS:2022fsw, ATLAS:2022uhq}, correlated across signal and backgrounds. Jet energy scale uncertainties range from 2-5\%~\cite{CMS:2016ucr, CMS:2015jsu}, leading to shape-based uncertainties of 1-3\%. A 10\% systematic uncertainty is added to account for calibration errors in signal and background predictions, uncorrelated across processes. Additionally, statistical uncertainties arise from the number of simulated events in specific L-GATr bins, ranging from 1-20\%, depending on the process and bin, and are uncorrelated across processes and bins. The total systematic uncertainty is about 30\%.

Figure \ref{fig:SS} shows the projected signal significance contours, as a function of $m(\W)$ and $g_\ell^{L, R}$ for SPM1 and SPM2, assuming $\mathcal{L}_\mathrm{int}=300$ fb$^{-1}$ (red) and 3000 fb$^{-1}$ (blue). The contours delimit the $5\sigma$ discovery region, $3\sigma$ excess contour, and the projected $95\%$ confidence level (CL) exclusion region if there is no statistically significant evidence of an excess. For SPM1 (SPM2) and $\mathcal{L}_\mathrm{int}=3000$ fb$^{-1}$, there is $5\sigma$ discovery potential for $m(\W)$ up to about 1.45 (0.45) TeV and $g_\ell^{L,R}=1$, and $m(\W) < 1.2$ (0.37) TeV for $g_\ell^{L,R} = 0.6$. The $3\sigma$ excess contour for SPM1 (SPM2) extends to $m(\W)\approx 3.6$ (1.9) TeV for $g_\ell^{L,R}=1$, but sensitivity decreases for lower $g_\ell^{L,R}$, reducing to about 2.75 (1) TeV for $g_\ell^{L,R} = 0.2$. Meanwhile, the expected $95\%$ CL exclusion bounds for SPM1 (SPM2) and $\mathcal{L}_\mathrm{int}=3000$ fb$^{-1}$ are $m(\W)< 4.45$ (2.3) TeV for $g_{\ell}^{L,R} = 1$, and $m(\W) < 3.9$ (1.9) TeV for $g_{\ell}^{L,R} = 0.6$.

In this Letter we investigate the production of a $\W$ boson at the LHC with negligible couplings to SM quarks, where the $\W$ interacts with SM weak bosons via triple gauge couplings and has a large decay width. This model could address discrepancies between experimental data and SM predictions and has not been previously considered at the LHC, and thus would have remained hidden from discovery. 

We introduce a pioneering search strategy that integrates point-cloud-based machine learning with the VBF topology, missing momentum, and final state leptons. Our results show that utilizing this strategy, along with a novel Lorentz-Equivariant machine learning classifier, greatly enhances the discovery potential and sensitivity of current and future LHC $\W$ searches. In quarkophobic $\W$ models with triple gauge couplings and large decay widths, the presence of $\W$ bosons decaying to a lepton and a neutrino can be probed for $\W$ masses up to 4.45 TeV, depending on the $\W$ boson coupling to SM leptons and weak bosons. We strongly encourage the ATLAS and CMS Collaborations to adopt this approach for upcoming experimental searches.

\begin{acknowledgments}
A. G. and U. S. Q. acknowledge funding from the Department of Physics and Astronomy at Vanderbilt University and the US National Science Foundation. 
This work is supported in part by NSF Awards PHY-2111554, PHY-1945366, PHY-2411502, and a Vanderbilt Seeding Success Grant. J. D. R.-\'A. gratefully acknowledges
the support of Universidad de Antioquia, the Colombian Science Ministry, and Sostenibilidad-UdeA.
\end{acknowledgments}

\appendix

\bibliography{apssamp}
\end{document}